\documentclass[useAMS,usenatbib]{mn2e}

 \usepackage{times}

\usepackage{amsmath}
\usepackage{amssymb}
\usepackage{graphicx}
\renewcommand{\d}{\mathrm{d}}
\newcommand{\dkl}{D_{\rmn{KL}}}
\renewcommand{\epsilon}{\varepsilon}
\newcommand{\evidence}{\mathcal{Z}}
\newcommand{\hMpc}{h\;\rmn{Mpc}^{-1}}

\renewcommand{\k}{{\bmath{k}}}
\newcommand{\likelihood}{L}
\newcommand{\Mpch}{\frac{\rmn{Mpc}}{h}}
\newcommand{\nbar}{\bar{n}}
\newcommand{\nbarinv}{\bar{n}^{-1}}
\newcommand{\PH}{P_H(\k)}
\newcommand{\Phat}{\widehat{P}(\k)}
\newcommand{\posterior}{\mathcal{P}}
\newcommand{\prior}{\Pi}

\newcommand{\x}{\bmath{x}}

\title[Inference from GC: The effect of the $P(\k)$ posterior]
{Cosmological parameter inference from galaxy clustering: The effect of the posterior distribution of the power spectrum}
\author[B. Kalus, W. J. Percival and L. Samushia]{B. Kalus$^{1}$\thanks{E-mail:
benedict.kalus@port.ac.uk}, W. J. Percival$^{1}$ and L. Samushia$^{2,1,3}$ 
\\
$^{1}$Institute of Cosmology \& Gravitation, Dennis Sciama Building, University of Portsmouth, Portsmouth, PO1 3FX, UK\\
$^{2}$Department of Physics, Kansas State University, 116, Cardwell Hall, Manhattan, KS, 66506, USA\\
$^{3}$National Abastumani Astrophysical Observatory, Ilia State University, 2A Kazbegi Ave., GE-1060 Tbilisi, Georgia}
\begin{document}

\date{Accepted . Received ; in original form \today}

\pagerange{\pageref{firstpage}--\pageref{lastpage}} \pubyear{2015}

\maketitle

\label{firstpage}

\begin{abstract}
  We consider the shape of the posterior distribution to be used when fitting cosmological models to power spectra measured from galaxy surveys. At very large scales,
  Gaussian posterior distributions in the power do not approximate the posterior distribution $\posterior_R$ we expect for a Gaussian
  density field $\delta_\k$, even if we vary the covariance matrix according to the model to be tested.
  We compare alternative posterior distributions with $\posterior_R$, both mode-by-mode and in terms of expected measurements of primordial non-Gaussianity parameterised by
  $f_\mathrm{NL}$. Marginalising over a Gaussian 
  posterior distribution $\posterior_f$ with fixed covariance matrix yields a posterior mean value of $f_\mathrm{NL}$ which, for a data set with the characteristics of 
  Euclid, will be underestimated by $\triangle f_\mathrm{NL}=0.4$, while for 
  the data release 9 (DR9) of the Sloan Digital Sky Survey (SDSS)-III Baryon Oscillation Spectroscopic Survey (BOSS) it will be underestimated by 
  $\triangle f_\mathrm{NL}=19.1$. 
  Adopting a different form of the posterior function means that we do 
  not necessarily require a different covariance matrix for each model to be tested: this dependence is absorbed into the functional form of the posterior. Thus, the computational burden of
  analysis is significantly reduced.
\end{abstract}


\section{Introduction}

Forthcoming galaxy surveys, such as the Dark Energy Spectroscopic Instrument \citep[DESI;][]{Schlegel:2011zz}, Euclid \citep{Laureijs:2011gra}\footnote{www.euclid-ec.org} 
and the Square Kilometre Array (SKA)
\footnote{www.skatelescope.org},
will constrain the Universe's expansion history, geometry and the growth of structure with unprecedented accuracy. The basic statistics containing large-scale structure information
are the 2-point clustering measurements, the correlation function and the galaxy power-spectrum $P(\k)$. 
As they form a Fourier pair, their information content
is the same and we focus only on the latter in this work. 
The linear galaxy power spectrum encodes a wealth of information about the physics of the Universe, allowing us to constrain 
cosmological models with baryon acoustic oscillations (BAO), gravitational models with redshift space distortions (RSD) and inflationary models with primordial non-Gaussianity, 
parametrised to first order by $f_\mathrm{NL}$. In order to do so, one has to know the likelihood and/or posterior
of power-spectra, which for simple cases can be calculated analytically. For general cases, one usually assumes the likelihood or posterior to be multi-variate Gaussian with a covariance matrix $C_{ij}\equiv \left\langle P(k_i) P(k_j)\right\rangle$. 
The estimation of the covariance matrix is a critical step in the analysis of data. 
Internal methods such as the sub-sample, jackknife and bootstrap methods have been widely used in the past, but 
\citet{Norberg:2008tg} have shown that they are not able to faithfully reproduce variances. Robust estimates are often instead obtained from mock galaxy catalogues.
In recent analyses of the Baryon Oscillation Spectroscopic Survey \citep[BOSS;][]{Dawson:2012va}, these were generated from 
second order Lagrangian Perturbation Theory matter fields using a friends-of-friends group finder \citep{Davis:1985} to find haloes 
\citep{Scoccimarro:2001cj, Manera:2012sc}. Their masses 
were calibrated by
comparisons with N-body simulations. A Halo Occupation Distribution then prescribed how to populate these haloes with mock galaxies, and the geometry and the efficiency of the
survey were sampled. Alternative methods for producing mocks include N-body simulations, comoving Lagrangian acceleration \citep*[COLA;][]{Tassev:2013pn} simulations or alternative simpler methods
such as pinpointing orbit-crossing collapsed hierarchical objects \citep*[PINOCCHIO;][]{Monaco:2001jg} or effective Zel'dovich approximation mocks \citep[EZmocks;][]{Chuang:2014vfa}. 
The covariance matrix is then the sample variance of the power spectra from the different mocks \citep{Manera:2012sc, Taylor:2012kz, Percival:2013sga}.
The covariance matrix computed from the mocks will depend on the cosmological model
that was used to generate them. It is computationally costly to produce mock catalogues for each possible cosmological model
and set of parameters to be tested,  so one usually chooses a cosmological model
which will produce a P(k) reasonably close to the measured one and uses the covariance 
matrix computed from the mocks created assuming that model. This approximation does not hold in general, 
especially at large scales.  In this article, we study other ways of approaching this problem, including using approximations to the true posterior distribution to obtain accurate inferences without requiring a covariance matrix for each cosmological model. We apply the most suitable of these approximation and the true distribution 
to provide a probability distribution function (PDF) for measurements of the non-Gaussianity parameter  $f_\mathrm{NL}$. Our result will provide a complementary method to analysing $f_\mathrm{NL}$ directly from $\delta(\x)$, as described in \citet{Verde:2013gv}.  

We proceed as follows. In Sec.~\ref{sec:toy}, we test the standard Gaussian posterior shapes mode-by-mode on a toy example where we measure the power spectrum itself. We do 
a similar test in Sec.~\ref{sec:BOSS} to study the impact of using different posterior distributions on a real survey, i.e. the Sloan Digital Sky Survey (SDSS)-III Baryon 
Oscillation Spectroscopic Survey (BOSS).  In Sec.~\ref{sec:optlike}, we
study alternative posterior shapes for power spectrum estimates inspired by CMB analyses. We test the most promising posterior distribution by postdicting a $f_\mathrm{NL}$-measurement
for a data sample like the data release 9 (DR9) of BOSS and we make predictions of Euclid $f_\mathrm{NL}$-measurements in Sec.~\ref{sec:fNL}. We conclude in Sec.~\ref{sec:conclusion}.

Throughout this work, we adopt a Bayesian framework and 
mark observed data with a hat, e.g. $\widehat P$ and quantities related to the hypothetical model with an $H$, e.g. $P_H$. We denote probability distributions with different letters $\likelihood$, $\posterior$, $\prior$ and $\evidence$, which we define in Tab.~\ref{tab:notation}, to make clear whether they depend on data and/or the model.
\begin{table}
 \centering
 \begin{minipage}{\linewidth}
  \caption{Notation used for probabilities.}
  \label{tab:notation}
  \begin{tabular}{@{}lll@{}}
  \hline
   symbol & name & 
   description \\\hline
   $\evidence(\widehat P)$ & evidence & probability of the data $\widehat P$\\
   $\likelihood(\widehat P\vert P_H)$ & likelihood & probability of the data $\widehat P$ given the hypothesis $P_H$\\
   $\posterior(P_H\vert \widehat P)$ & posterior & probability of the hypothesis $P_H$ given the data $\widehat P$\\
   $\prior(P_H)$ & prior & probability of the hypothesis\\
\hline
\end{tabular}
\end{minipage}
\end{table}
The ubiquitous Bayesian equation thus reads for this example
\begin{equation}
 \posterior\left(P_H\left\vert \widehat P\right.\right)
 =\frac{\likelihood\left.\left(\widehat P\right\vert P_H\right)\prior\left(P_H\right)}{\evidence\left(\widehat P\right)}.
  \label{eq:Bayes}
\end{equation}

\section{The Power Spectrum Likelihood}
\label{sec:toy}
In this section, we elaborate the analytic likelihood and posterior functions of the galaxy clustering power spectrum assuming a Gaussian density field. We consider this posterior function as the "truth" and compare it to commonly used approximations of the galaxy power spectrum posterior function for single modes, which we shall introduce in Sec.~\ref{sec:CommonApprox}.

\subsection{The True Distribution of $\widehat{\vert\delta_\k\vert}$ Under the Assumption of a Gaussian Density Field}
\label{sec:deltakL}
The positions of the galaxies in a survey can be transformed into a galaxy over-density field 
\begin{equation}
 \delta(\bmath{x})\equiv\frac{n(\bmath{x})-\bar{n}(\bmath x)}{\bar{n}(\bmath x)},
\end{equation}
where $n(\x)$ is the measured galaxy number density and $\bar n(\bmath x)$ the expected value.
Fourier transforming $\delta(\bmath{x})$ yields
\begin{equation}
 \delta_{\bmath{k}}\equiv\frac{1}{V} \int\d^3\bmath{x}\delta(\bmath{x})\exp(i\bmath{kx})
\end{equation}
whose covariance matrix 
\begin{equation}
 \left\langle\delta_{\k_1}\delta_{\k_2}^\ast\right\rangle=\frac{(2\pi)^3}{V}\delta_D(\bmath{k_1}-\bmath{k_2})P(\bmath{k_1})
 \label{eq:Pasdeltacov}
\end{equation}
is given by the power spectrum $P(\bmath{k})$.
Following the standard assumption that $\delta_\k$ forms a Gaussian random field, the probability of measuring a particular value of the real and imaginary parts 
$\left(\widehat{\delta_u}, \widehat{\delta_v}\right)$ 
of a single 
$\widehat{\delta_{\bmath{k}}}=\widehat{\delta_u}+i\widehat{\delta_v}$ is a zero centred Gaussian distribution with standard deviation half the true power $\frac{1}{2}P_T(\k)$:
\begin{align}
 \evidence\left(\widehat{\delta_{\bmath u}}\right)=\frac{1}{\sqrt{\pi P_T(\bmath{k})}}\exp
  \left(-\frac{\widehat{\delta_u}^2}{P_T(\bmath{k})}\right),\nonumber\\
 \evidence\left(\widehat{\delta_{\bmath v}}\right)=\frac{1}{\sqrt{\pi P_T(\bmath{k})}}\exp
  \left(-\frac{\widehat{\delta_v}^2}{P_T(\bmath{k})}\right).
 \label{eq:deltaL} 
\end{align}
We use the letter $\evidence$ here, because we have assumed that the true power is known, i.e. the distribution only depends on the data (c.f. Tab.~\ref{tab:notation}).
The distribution of the absolute value $\widehat{\vert\delta_{\bmath k}\vert}=\sqrt{\widehat{\delta_{\bmath u}}^2+\widehat{\delta_{\bmath v}}^2}$ is given by a Rayleigh distribution:
\begin{align}
  \evidence_R\left(\widehat{\vert\delta_{\bmath k}\vert}\right)=&
  \int\d\widehat{\delta_{\bmath u}}\int\d\widehat{\delta_{\bmath v}}\evidence\left(\widehat{\delta_{\bmath u}}\right)\evidence\left(\widehat{\delta_{\bmath v}}\right)
  \delta_D\left(\widehat{\vert\delta_{\bmath k}\vert}-\sqrt{\widehat{\delta_{\bmath u}}^2+\widehat{\delta_{\bmath v}}^2}\right)\nonumber\\
  =&\frac{2\widehat{\vert\delta_{\bmath{k}}\vert}}{P_T(\bmath{k})}\exp\left(-\frac{\widehat{\vert\delta_{\bmath{k}}\vert}^2}{P_T(\bmath{k})}\right).
 \label{eq:RayleighZ} 
\end{align}
Throughout this article, we regard Eq.~\eqref{eq:RayleighZ} as the "true" distribution of $\widehat{\vert\delta_\k\vert}$ to which we compare several approximations later.

Any model dependence enters the Rayleigh distribution only in the covariance of the density field, which is equal to the true power spectrum. The position of the distribution's peak equals the value of the true power.
Measurements of $\widehat{\delta_{\bmath{k}}}$ have been used to make cosmological inferences when they have been 
further decomposed into spherical harmonics and spherical Bessel functions, because radial and angular modes can be
distinguished, allowing an easy analysis of redshift-space distortions. However, this method is rather complex and computationally expensive \citep{Heavens:1994iq, Percival:2004fs}. It is difficult to linearly compress $\widehat{\delta_\k}$ efficiently maximally retaining information.

\subsection{The Posterior in Terms of the Power}
\label{sec:PRrewr}
We can rewrite the Rayleigh distribution in terms of the power.  We replace $P_T(\k)$ with $P_H(\k)$, and $\widehat{\delta_\k}$ with $\sqrt{\Phat}$ in equation (\ref{eq:RayleighZ}) which in this way depends on both data and model, and hence becomes a likelihood (c.f. Tab.~\ref{tab:notation}):
\begin{align}
  \likelihood_R\left.\left(\widehat P(\bmath{k})\right\vert P_{H}(\bmath{k})\right)
  =\frac{2\sqrt{\Phat}}{P_H(\bmath{k})}\exp\left(-\frac{\Phat}{P_H(\bmath{k})}\right).
 \label{eq:RayleighL} 
\end{align}
We can use Bayes' theorem (cf. Eq.~\eqref{eq:Bayes}) to find the posterior.
It is standard to assume a uniform prior
\begin{equation}
 \prior\left(P_{H}(\bmath{k})\right)=\begin{cases}
                                      \frac{1}{P_{\rmn{max}(\k)}}&\text{, if }0\leq P_H(\k) \leq P_{\rmn{max}}(\k),\\
				      0&\text{otherwise,}
                                     \end{cases}
 \label{eq:uniformPHprior}
\end{equation}
which requires an arbitrary choice of $P_{\rmn{max}}(\k)$. We assume that $P_{\rmn{max}}(\k)$ is far in the right tail of the likelihood such that 
$\frac{\prior\left(P_{H}(\bmath{k})\right)}{\evidence\left(\widehat P(\k)\right)}$ is effectively constant and hence acts only as a normalisation factor. Thus, for the "true" posterior
we have
\begin{align}
 \posterior_R\left(P_{H}(\bmath{k})\left\vert \widehat P(\bmath{k})\right.\right)&= 
 \frac{L_R\left.\left(\widehat P(\bmath{k})\right\vert P_{H}(\bmath{k})\right)}{\int\d P_H~ L_R\left.\left(\widehat P(\bmath{k})\right\vert P_{H}(\bmath{k})\right)}\nonumber\\
 &\propto\frac{2\sqrt{\Phat}}{P_H(\bmath{k})}\exp\left(-\frac{\Phat}{P_H(\bmath{k})}\right).
 \label{eq:PRL}
\end{align}
As $\widehat P$ is a constant in the posterior, one can rewrite Eq.~\eqref{eq:PRL} such that the log-posterior only depends on the ratio $\widehat P(\k)/P_H(\k)$: 
\begin{equation}
	-2\ln(\mathcal P_R)=2M\ln\left(\frac{P_H(\k)}{\widehat P(\k)}\right)+2M\frac{\widehat{P}(\k)}{P_H(\k)}+\text{const}.
	\label{eq:logpost}
\end{equation}
Then we follow the method of \citet{Hamimeche:2008ai} and 
introduce
\begin{equation}
	\gamma(x)\equiv\sqrt{-\ln(x)+x}
\end{equation}
to make Eq.~\eqref{eq:logpost} look more quadratic:
\begin{equation}
	-2\ln(\mathcal P_R)=2M\left[\gamma\left(\frac{\widehat{P}(\k)}{P_H(\k)}\right)\right]^2+\text{const}.
\end{equation}
We can also define 
\begin{equation}
	P_\gamma(\k)\equiv P_f(\k)\gamma\left(\frac{\widehat{P}(\k)}{P_H(\k)}\right)
\end{equation}
for some fiducial model with power $P_f$. $P_\gamma$ has then a symmetric Gaussian posterior with a fixed variance $\tilde C_k=\frac{2P_f^2(\k)}{M}$ evaluated for our fiducial model:
\begin{equation}
	-2\ln(\mathcal P_R)=4 P_\gamma\tilde C_k^{-1}P_\gamma+\text{const}.
	\label{eq:PHL}
\end{equation}

In general. things are more complicated than this simple picture. For example, the survey geometry leads to a convolution of $\delta_\k$, and non-linear effects distort the small scale mode distribution. Ideally, we would like to use a single distribution, and this should be matched to simulations \citep[e.g.][]{Blot:2014pga}. In order to broaden the choice, we  also consider a number of forms for the likelihood inspired by CMB analyses.

\subsection{Common Approximations of the Likelihood/Posterior of the Power Spectrum}
\label{sec:CommonApprox}

Often, the power-spectrum is directly analysed, incorrectly assuming it follows a Gaussian distribution, thus the distribution of a finite empirical realisation of the power spectrum $\Phat$ would read 
\begin{equation}
 \evidence\left(\widehat P(\bmath{k})\right)=\frac{\exp\left(-\frac{1}{2}\frac{\left[\widehat P(\bmath{k})-P_{T}(\bmath{k})\right]^2}{C_{\bmath{k}}}\right)}{\sqrt{2\pi C_{\bmath{k}}}},
 \label{eq:PGauss}
\end{equation}
where $C_{\bmath{k}}\equiv\left\langle P_T^2(\bmath k)\right\rangle=\frac{2P_T^2(\k)}{M}$ is the variance of the true power spectrum $P_T$ at 
a bin centred around $\bmath{k}$ comprising $M$ independent modes. Note that we assume that the widths and positions of the $\k$-bins are such that window effects are negligible \citep{Feldman:1993ky} and different modes are independent.

As in Sec.~\ref{sec:PRrewr}, we replace $P_T(\k)$ with $P_H(\k)$ in equation (\ref{eq:PGauss}) making it a likelihood (c.f. Tab.~\ref{tab:notation}):
\begin{equation}
 \likelihood\left.\left(\widehat P(\bmath{k})\right\vert P_{H}(\bmath{k})\right)
 =\frac{\exp\left(-\frac{1}{2}\frac{\left[\widehat P(\bmath{k})-P_{H}(\bmath{k})\right]^2}{C_{\bmath{k}}^H}\right)}{\sqrt{2\pi C_{\bmath{k}}^H}},
 \label{eq:PvaryL}
\end{equation}
where $C_{\bmath{k}}^H\equiv\left\langle P_H^2(\bmath k)\right\rangle$ is the variance for the hypothetical power spectrum $P_H(\bmath k)$.

However, in practice one chooses a fiducial model with power spectrum $\tilde P(\k)$ and estimates the 
variance $\tilde C_\k\equiv\left\langle\tilde P^2(\bmath k)\right\rangle$ for this particular choice:
\begin{equation}
 \likelihood\left.\left(\widehat P(\bmath{k})\right\vert P_{H}(\bmath{k}),\tilde C_{\bmath{k}}\right)= 
  \frac{\exp\left(-\frac{1}{2}\frac{\left[\widehat P(\bmath{k})-P_{H}(\bmath{k})\right]^2}{\tilde C_{\bmath{k}}}\right)}{\sqrt{2\pi\tilde C_{\bmath{k}}}}. 
 \label{eq:LfixedL}
\end{equation}
For mock based variance calculations, $\tilde P(\k)$ is the cosmology of the mocks used in their analysis.

We can again use Bayes' theorem (cf. Eq.~\ref{eq:Bayes}) and assume the same uniform prior as before to find the posterior. For the posterior
assuming a Gaussian distribution in $\Phat$ with model-dependent covariance we have
\begin{align}
 \posterior_D\left(P_{H}(\bmath{k})\left\vert \widehat P(\bmath{k})\right.\right)&= 
 \frac{L\left.\left(\widehat P(\bmath{k})\right\vert P_{H}(\bmath{k})\right)}{\int\d P_H~ L\left.\left(\widehat P(\bmath{k})\right\vert P_{H}(\bmath{k})\right)}\nonumber\\
 &\propto\frac{\exp\left(-\frac{1}{2}\frac{\left[\widehat P(\bmath{k})-P_{H}(\bmath{k})\right]^2}{C_{\bmath{k}}^H}\right)}{\sqrt{2\pi C_{\bmath{k}}^H}},
 \label{eq:PMvaryL}
\end{align}
where we adopt the subscript notation $\posterior_D$ of \citet{Hamimeche:2008ai}.
Note that both the 
exponential and the covariance matrix $C_{\bmath{k}}^H$ depend on $\PH$.

If a fixed covariance is assumed, we have to apply the Bayesian Eq.~\eqref{eq:Bayes} to Eq.~\eqref{eq:LfixedL} giving
\begin{equation}
 \posterior_f\left(P_{H}(\bmath{k})\left\vert \widehat P(\bmath{k}),\tilde C_{\bmath{k}}\right.\right)\propto 
  \frac{\exp\left(-\frac{1}{2}\frac{\left[\widehat P(\bmath{k})-P_{H}(\bmath{k})\right]^2}{\tilde C_{\bmath{k}}}\right)}{\sqrt{2\pi\tilde C_{\bmath{k}}}}. 
 \label{eq:PMfixedL}
\end{equation}

\subsection{A Simple Test of the Posterior Shapes for the isotropically averaged power spectrum}

In this subsection we combine the single mode posterior functions to posterior functions of the band-power. We do not take any anisotropic effects, such as redshift space distortions, 
into account. This is conservative because the effective volume for higher multipole moments 
(cf. Eq. \ref{eq:Veff}) is smaller, therefore containing fewer independent modes and thence amplifying the effect of choosing different posterior shapes.

In Gaussian cases, we suppose that our volume is large enough to accommodate $M$ independent complex Gaussian distributed samples of $\delta_\k$ such that we can use
\begin{equation}
 C_{ab}=\frac{2}{M}\delta_D(\k_a-\k_b)P^2({\bf k}_a).
  \label{eq:cov}
\end{equation}
to calculate the covariance matrices at higher numbers of modes $M$. We can obtain the band-power version of 
$\posterior_R\left(P_{H}(\bmath{k})\left\vert\widehat{\vert\delta_{\bmath k}\vert}\right.\right)$ by multiplying together the single mode expressions.

\begin{figure}
 \centering
 \includegraphics[width=\linewidth]{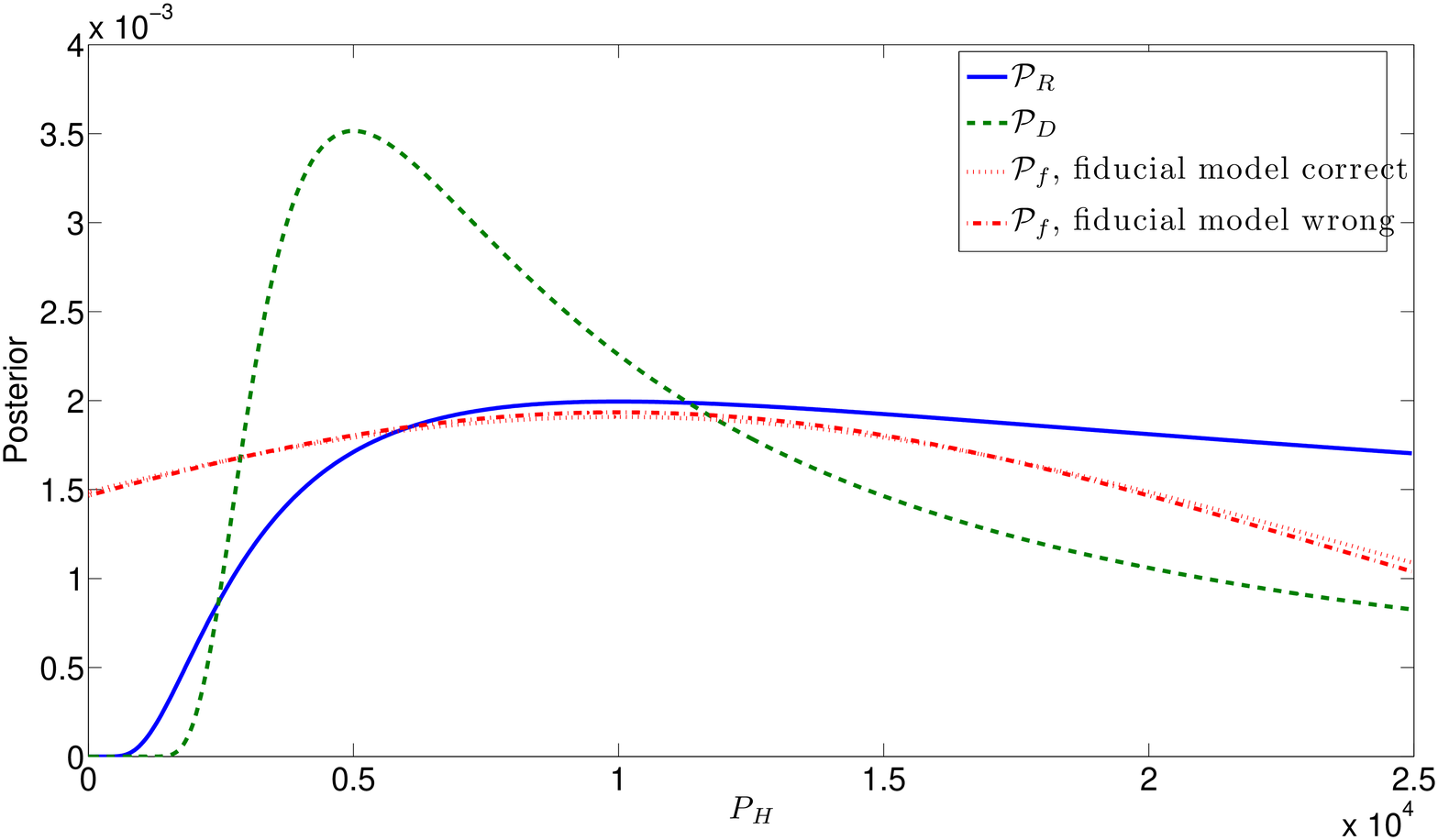}
 \includegraphics[width=\linewidth]{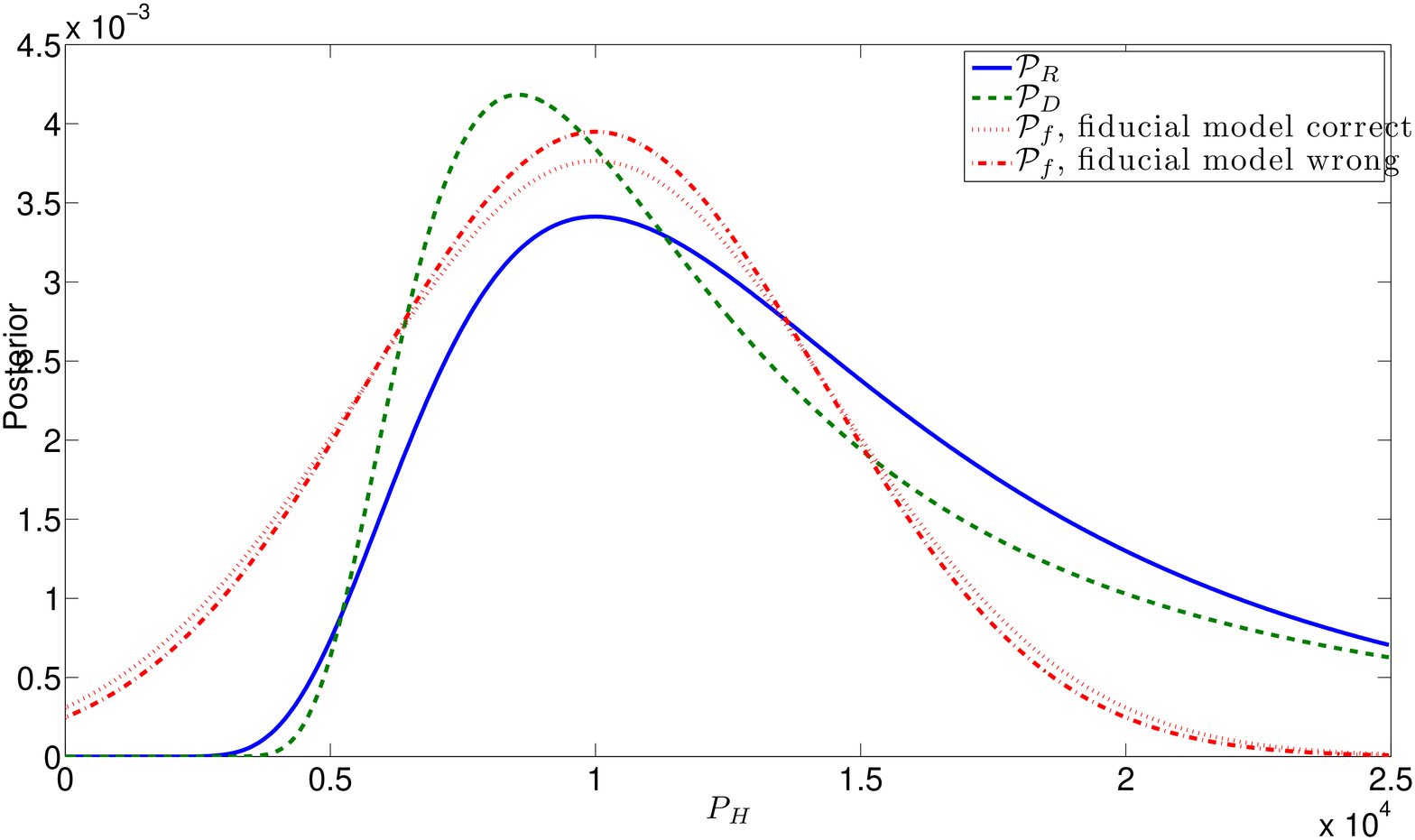}
 \includegraphics[width=\linewidth]{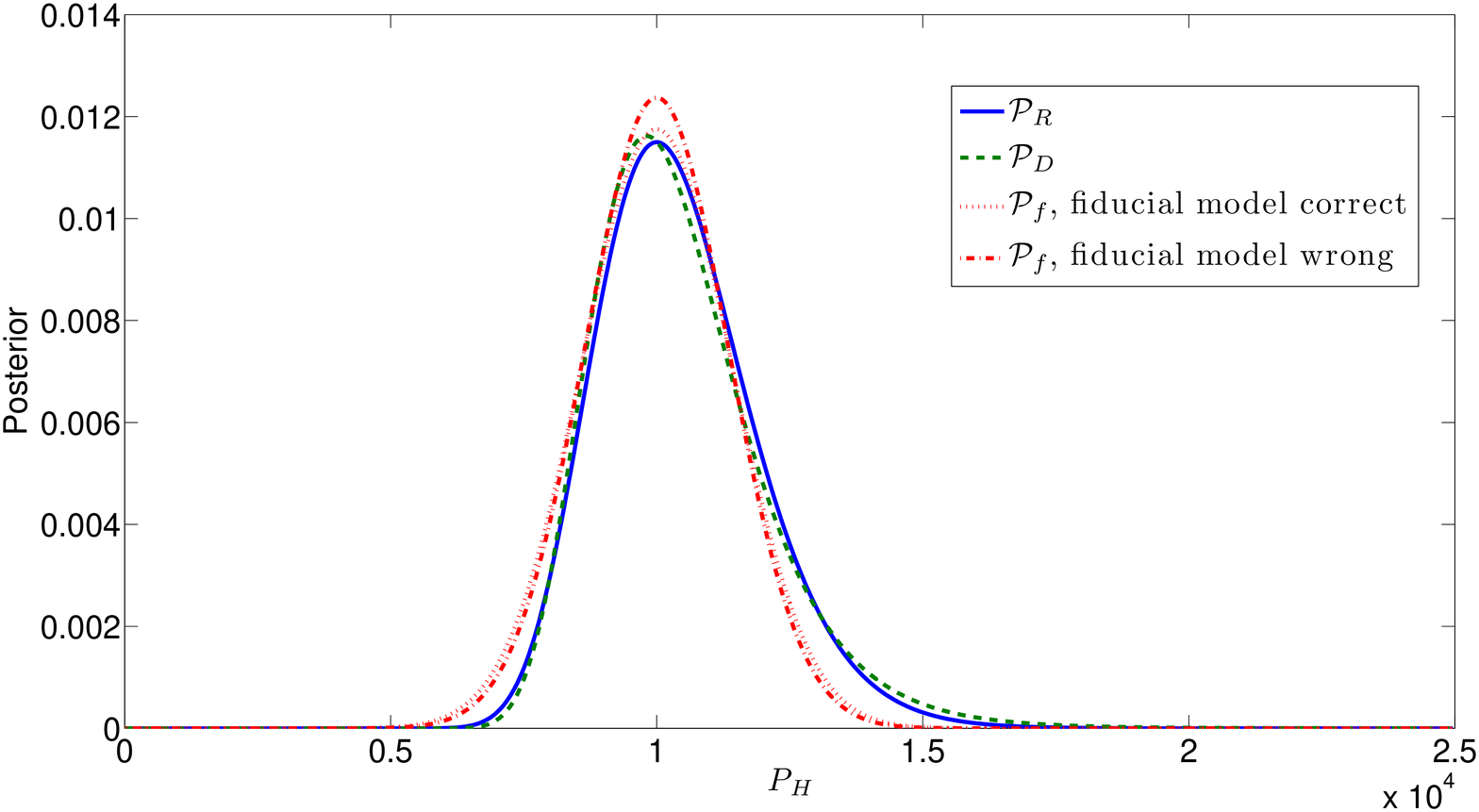}
 \includegraphics[width=\linewidth]{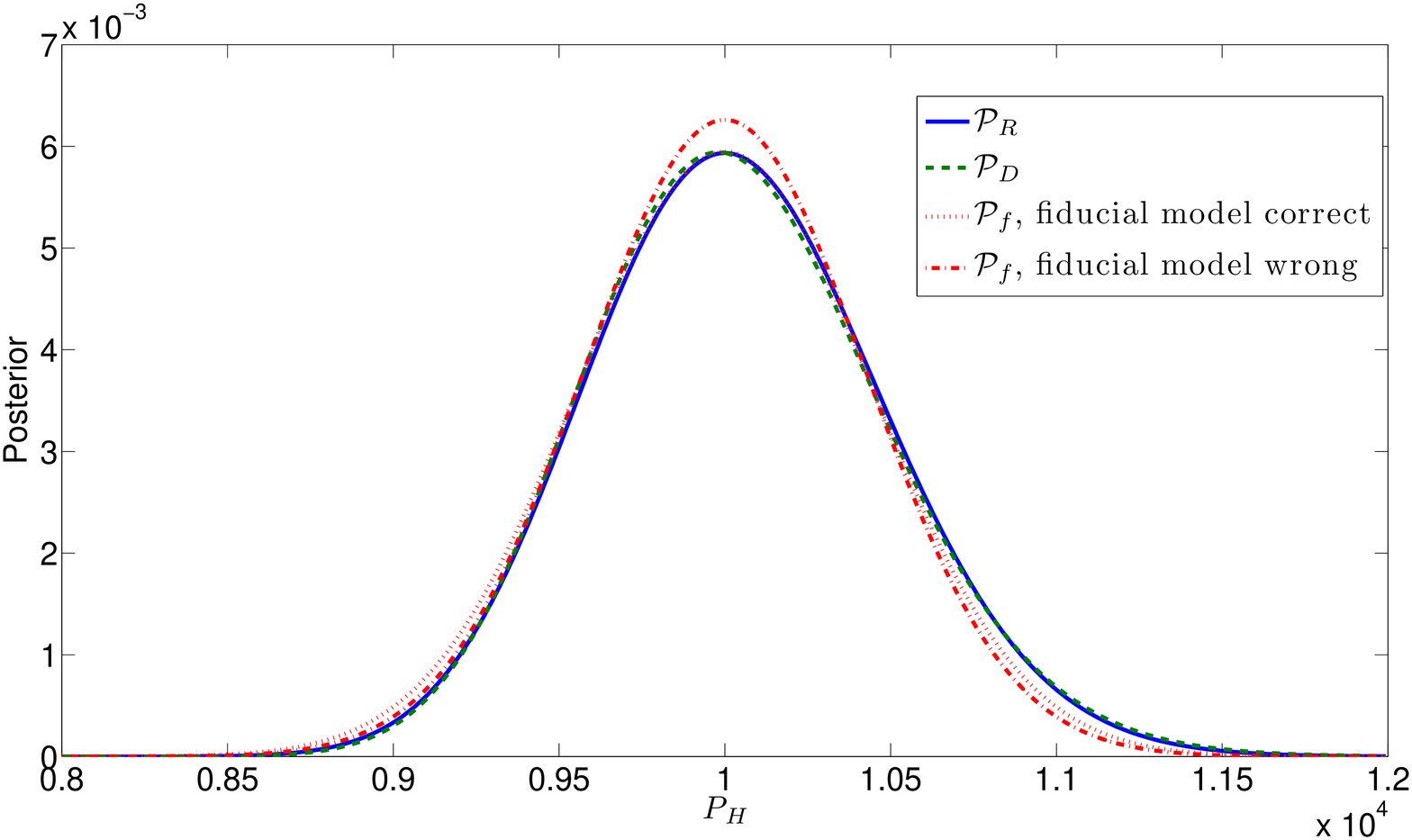}
 \caption{Comparison of different posterior distribution functions for 1, 10, 100 and 1000 independent modes (from top to bottom). 
  The blue line represents the product of
  single Rayleigh distributed modes (true posterior distribution) and some of the approximations, such as the Gaussian posterior distribution 
  with a model-dependent covariance (green), and the 
  Gaussian posterior where the covariance is estimated for a fixed fiducial model (red). The posterior takes the form of the dotted red line if the fiducial and 
  the true power spectra agree, the dashed-dotted line shows the effect of choosing a fiducial model of which the power spectrum is wrong by 5 per cent.}
 \label{fig:Lcomp}
\end{figure}

The three different posterior shapes of $P_H$ are plotted in Fig.~\ref{fig:Lcomp}. In the top panel of Fig.~\ref{fig:Lcomp}, we plot single mode posterior distributions for which we
adopt $\widehat{\left\vert\delta_\k\right\vert}=100$ and $P_T(\k)=\widehat{P}(\k)=\widehat{\left\vert\delta_\k\right\vert}^2=10000$. Note that a different choice would 
shift the peak 
positions and normalisation factor, but preserve the shapes. We make two different choices for the fixed covariance to
see the effect of making the wrong assumption. For the dotted red line, we choose the covariance matrix which corresponds to the true power spectrum 
$P_T(\k)$, i.e. $\tilde{C}_\k=2P_T^2(\k)=50000000$, and for the dashed-dotted line, we consider that our guess of the power spectrum is 5 per cent lower than the actual 
power spectrum, i.e. $\tilde{C}_\k=45125000$. The panels in the middle and at the bottom of Fig.~\ref{fig:Lcomp} show the posterior distributions for 10 and 100 independent modes
respectively. 

\begin{figure}
 \centering
 \includegraphics[width=\linewidth]{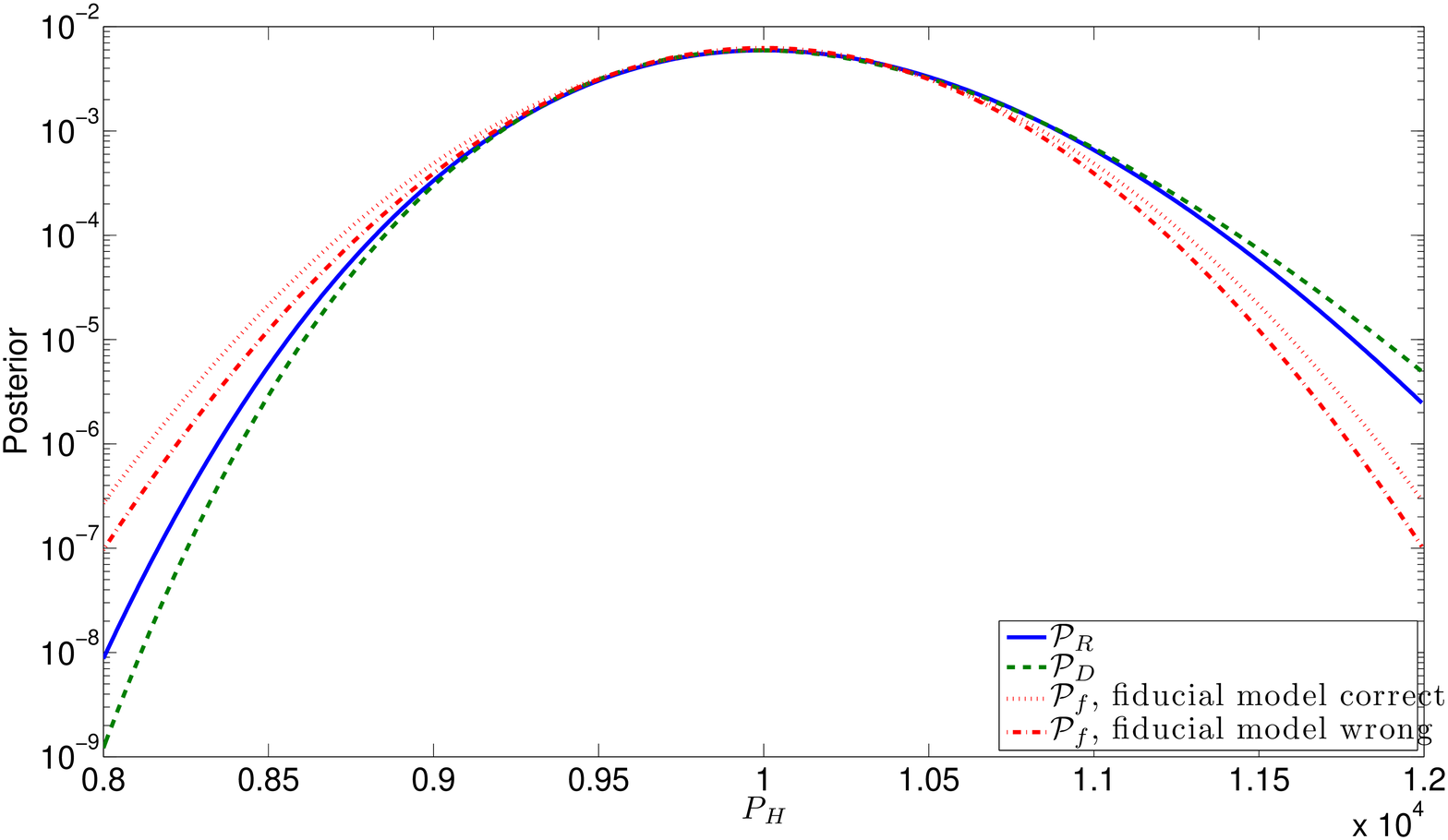}
 \caption{Same as the bottom panel of Fig.~\ref{fig:Lcomp}, but with a logarithmic ordinate.}
 \label{fig:Lcomplog}
\end{figure}

Fig.~\ref{fig:Lcomp} and \ref{fig:Lcomplog} show that different choices of the covariance matrix provide very different posterior distributions 
for a small number of modes, but if we can increase the number of
independent modes, we see the effect of the central limit theorem and the posterior distribution functions 
become more and more similar. We observe that the maximum of the fixed-covariance 
posterior always agrees with the true value, even if the wrong fiducial model has been chosen. However, if we choose the wrong covariance matrix, we over- or underestimate the error
of our measurements. If we do not fix the covariance, the best fit, i.e. the maximum of the posterior, has an offset with regard to 
the true value, which decreases as the number of modes increases. We also notice the long right tails of the varying-covariance Gaussian and the posterior measured from 
$\widehat{\left\vert\delta_\k\right\vert}$. The logarithmic plot in Fig.~\ref{fig:Lcomplog}  shows that the tails of all approximations disagree with the true posterior distribution. However, $\posterior_D$ is closest to the truth.

\subsection{Application to a Real Survey}
\label{sec:BOSS}

We have seen that a Gaussian distribution for $P_H(\k)$ is not a good approximation to the true Rayleigh distribution if the number of modes is small. In this section,
we study whether this has an impact on a real survey. We will base our analysis on an analytic linear error for the power spectrum and errors, but use survey parameters 
for the data release 11 (DR11) of BOSS. 
For a real survey, we have to take into account that the 
discrete positions of the galaxies in a given survey are sampled from a continuous random field by a Poisson point process \citep{Feldman:1993ky}. To take this sampling process into
account,Eq.~\eqref{eq:Pasdeltacov} becomes
\begin{equation}
 \left\langle\delta_{\k_1}\delta_{\k_2}^\ast\right\rangle=\frac{(2\pi)^3}{V}\delta_D(\bmath{k_1}-\bmath{k_2})\left[P(\bmath{k_1})+\nbarinv\right],
 \label{eq:Pnasdeltacov}
\end{equation}
and hence also 
\begin{equation}
 \posterior\left(P_{H}(\bmath{k})\left\vert\widehat{\vert\delta_{\bmath k}\vert}\right.\right) 
  \propto\frac{\widehat{\vert\delta_{\bmath{k}}\vert}\exp\left(-\frac{\widehat{\vert\delta_{\bmath{k}}\vert}^2}{P_H(\bmath{k})+\nbar^{-1}}\right)}
  {P_H(\bmath{k})+\nbar^{-1}}.
 \label{eq:PMRayleighSN}
\end{equation}
The average number density $\nbar=2\times 10^{-4}\;\frac{h^3}{\rmn{Mpc}^3}$ can be calculated from the number of galaxies contained in the BOSS DR11 CMASS sample 
(690,826) and its survey volume $V_S=10\;\rmn{Gpc}^3$ 
\citep{Anderson:2013zyy} assuming $h=0.7$. For the covariance matrices of the Gaussians, we need to know the number of modes \citep{Feldman:1993ky,Tegmark:1997rp}
\begin{equation}
 M=V_nV_{\rmn{eff}}(\k),
\end{equation}
where 
\begin{equation}
 V_n\equiv \frac{k_n^2 \triangle k_n}{2\pi^2}
 \label{eq:Vn}
\end{equation}
is the $k$-space-``volume'' of the $n$th $k$-bin centred at $k_n$ with width $\triangle k_n$, and 
\begin{equation}
 V_{\rmn{eff}}(\k)\equiv V_S\left[\frac{\nbar P(\k)}{1+\nbar P(\k)}\right]^2
 \label{eq:Veff}
\end{equation}
is the effective volume.
\citet{Anderson:2013zyy} calculate the power spectrum in Fourier modes averaged over bin widths of 
$\triangle k=0.008 h\;\rmn{Mpc}^{-1}$. The values of the $k$-bin centres and their corresponding number of modes $M$ are M=18, 180 and 500 in the three lowest $k$-bins centred at $k=0.004, 0.012$ and $0.02\;\mathrm{Mpc}\;h^{-1}$. We model the measured power
spectrum as $\widehat{P}(\k)=b^2P_{\rmn{lin}}(\k)$, where $b=1.87$ is the large-scale bias and $P_{\rmn{lin}}(\k)$ is a linear power spectrum produced by CAMB \citep{Lewis:1999bs}. 
For the other measurement we take $\widehat{\left\vert\delta_\k\right\vert}=\sqrt{\widehat P(\k)+\nbar^{-1}}$. The resulting posterior distributions for the three lowest
$k$-bins are plotted in Fig.~\ref{fig:OLN}. At the largest scales, i.e. $k=0.004h\;\rmn{Mpc}^{-1}$, neither $\posterior_D$ or $\posterior_f$ match $\posterior_R$. 
At $k=0.012\hMpc$ and $k=0.02\hMpc$ $\posterior_f$ and $\posterior_D$ become more similar, but neither of them features the asymmetric shape of 
$\posterior_R$. Additionally, $\posterior_f$ and $\posterior_D$ produce smaller error bars compared to $\posterior_R$.
We can also numerically compare the distributions if we introduce the Kullback-Leibler (KL) divergence \citep{Kullback:1951}. A distribution $\posterior_1$ is 
``better'' than $\posterior_2$, 
if the loss of information due to approximating the true distribution with $\posterior_1$ is less than the same loss caused by using $\posterior_2$ as an approximation.
If we use a probability density function (pdf) $g$ to approximate another pdf $f$, 
a measure of the loss of information is given by the KL divergence 
\begin{equation}
 \dkl\left(g\vert\vert f\right)\equiv\int_{-\infty}^{\infty}\d x f(x)\ln\left(\frac{f(x)}{g(x)}\right).
\end{equation}
The KL divergences given in Tab.~\ref{tab:DKLBOSS} tell us the same story as Fig.~\ref{fig:OLN}. The KL divergences of the Gaussian approximation with a 
varying covariance $\posterior_D$ is at all scales less than the KL divergence of $\posterior_f$, i.e. $\posterior_D$ is a better approximation to the 
true $\posterior_R$. On the downside, its best fit has an offset with respect to $\posterior_R$. We will therefore investigate alternative posterior 
shapes in the next section.

\begin{table}
 \centering
 \begin{minipage}{\linewidth}
  \caption{Kullback-Leibler divergences of the different approximations with respect to the true $\posterior_R$ at different scales $k_n$ for BOSS DR11 CMASS.}
  \label{tab:DKLBOSS}
  \begin{tabular}{@{}llll@{}}
  \hline
   \tiny $k_n\;\frac{\rmn{Mpc}}{h}$ & \tiny $\dkl\left(\posterior_D||\posterior_R\right)$ & 
   \tiny $\dkl\left(\posterior_f||\posterior_R\right)$ & \tiny
    $\dkl\left(\posterior_f^{\rmn{wrong }}||\posterior_R\right)$ \\
 \hline
  0.004	& 0.0213221 & 0.382926 & 0.335588 \\
  0.012	& 0.00451188 & 0.0341685 & 0.0374833 \\
  0.02	& 0.00336263 & 0.0138535 & 0.0199032 \\
\hline
\end{tabular}
\end{minipage}
\end{table}

\section{Studying Alternative Posterior Shapes}
\label{sec:optlike}

We have seen in the previous sections that the true posterior distribution $\posterior_R$ is not well approximated by either $\posterior_f$ or $\posterior_D$ if the number of
independent modes is low, which is the case at large scales, i.e. small values of $k$. A similar problem arises when cosmological models are fitted to cosmic
microwave background (CMB) power spectra, which are Wishart distributed. \citet*{Bond:1998qg}, \citet*{Smith:2005ue}, 
\citet{Percival:2006ss} and \citet{Hamimeche:2008ai} have studied alternative distribution
shapes that approximate the Wishart distribution. 
We take a similar approach to
\citet{Verde:2003ey} and \citet{Percival:2006ss} and expand the natural logarithm of Eq.~\eqref{eq:PRL} around the maximum $\PH\equiv(1+\epsilon)\widehat{\left\vert\delta_\k
\right\vert}^2$:\footnote{For realistic, noisy measurements of $\widehat{\vert\delta_\k\vert}$ and $\widehat P(\k)$, $P_H(\k)$ has to be replaced by 
$P_H(\k)+\nbarinv$ everywhere in this section. For simplicity, we do not write the noise explicitly.}
\begin{equation}
 -2\ln\left(\posterior_R\right)=2M\left(\frac{\epsilon^2}{2}-\frac{2\epsilon^3}{3}+\frac{3\epsilon^4}{4}+\mathcal{O}\left(\epsilon^5\right)\right)+\text{const.}
 \label{eq:PRTaylor}
\end{equation}
This equation agrees to third order with the Taylor expansions of the logarithms of the following distributions:
\begin{itemize}
 \item the inverse cubic normal (ICN) distribution \citep{Smith:2005ue}
    \begin{equation}
      -2\ln(\posterior_{\rmn{ICN}})=18\tilde C^{-1}_\k\left[\Phat-\Phat^{4/3}\PH^{-1/3}\right]^2,
      \label{eq:PICN}
    \end{equation}
 \item the offset log-normal (OLN) distribution
    \begin{equation}
      -2\ln(\posterior_{\rmn{OLN}})=2(1+a)\tilde C^{-1}_\k\left[\Phat\ln\left(\frac{\PH+a\Phat}{\Phat+a\Phat}\right)\right]^2
    \end{equation}
    if $a=-1/4$, 
 \item and combinations of any of the distributions given in chapter 5.1 of \citet{Percival:2006ss}.
\end{itemize}

\begin{figure}
 \centering
 \includegraphics[width=\linewidth]{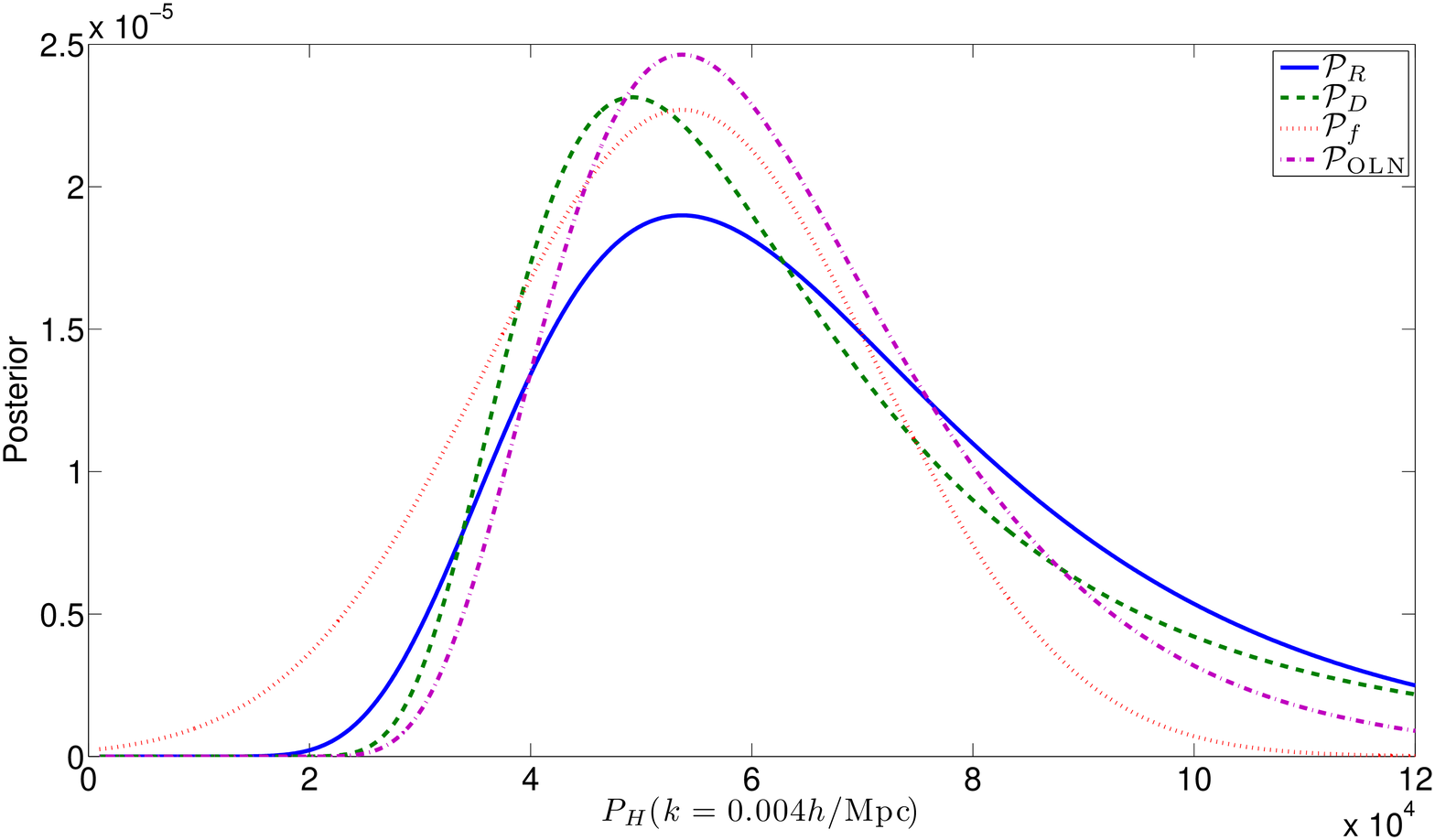}
 \includegraphics[width=\linewidth]{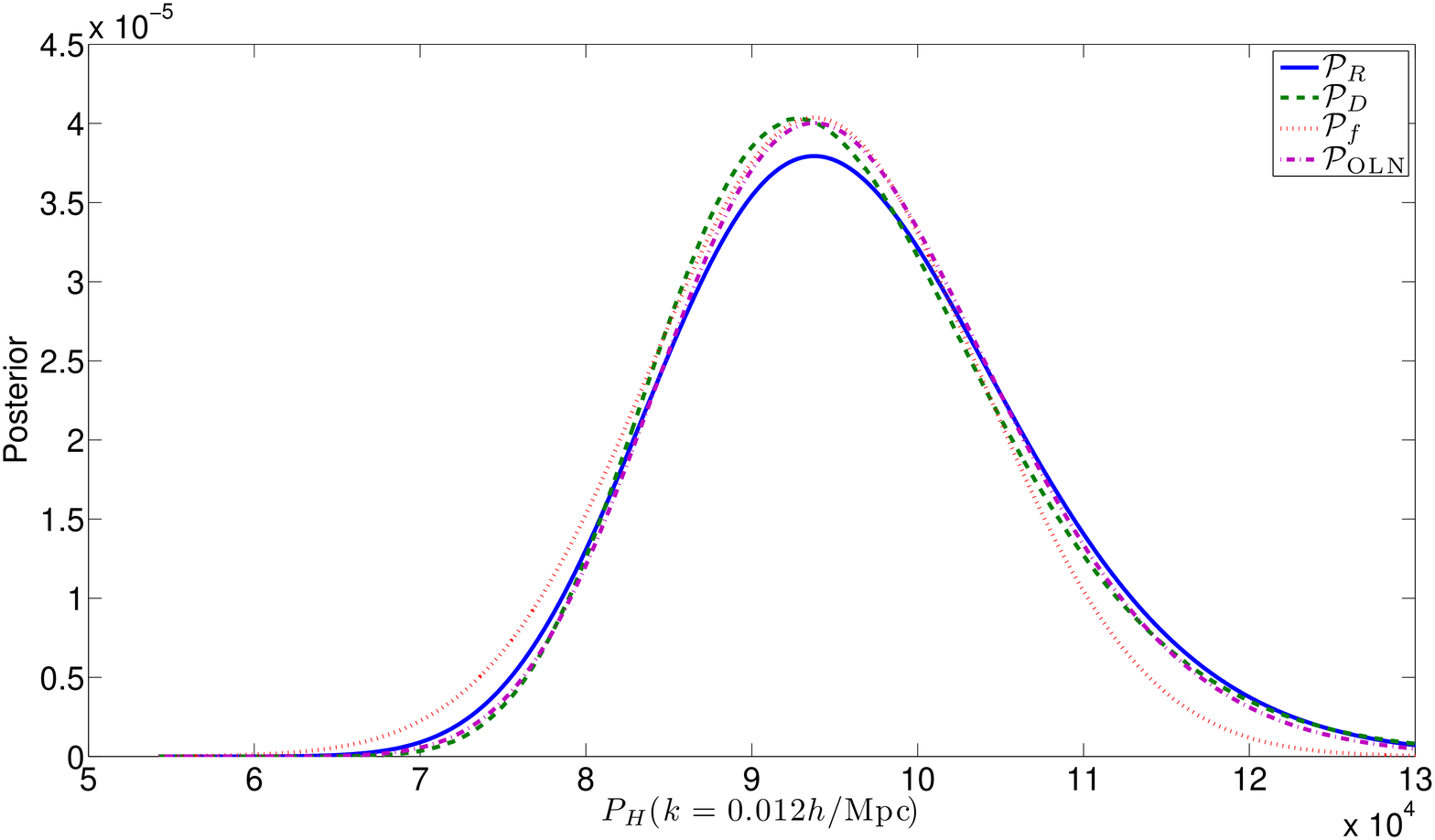}
 \includegraphics[width=\linewidth]{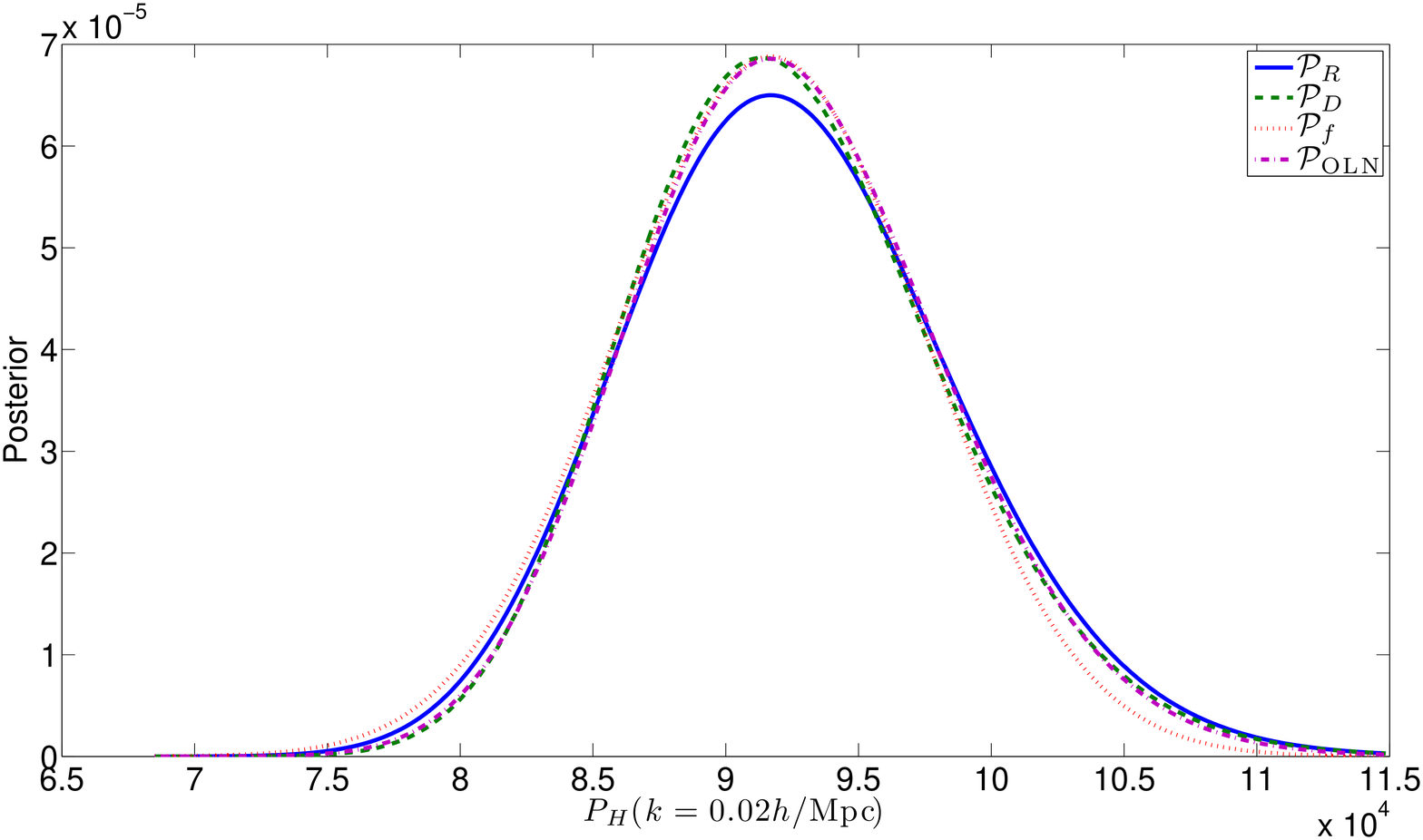}
 \caption{Posterior distribution functions of the hypothetical power spectrum $P_H(\k)$ for the three lowest $\k$-bins of BOSS DR11 CMASS. The colour coding is the same as in 
 Fig.~\ref{fig:Lcomp}, with the addition of the offset log-normal (OLN) posterior distribution plotted in magenta.}
 \label{fig:OLN}
\end{figure}

\begin{figure}
 \centering
 \includegraphics[width=\linewidth]{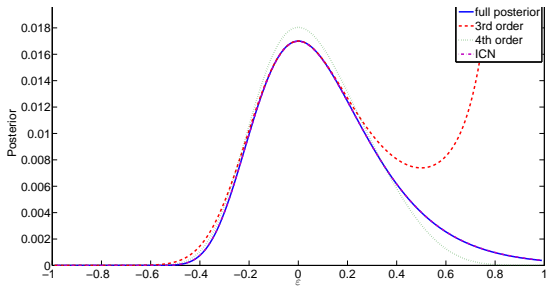}
 \caption{Third and fourth order Taylor expansion to the true posterior shape $\posterior_R$ with $M=20$ modes. The x-axis is a perturbation $\varepsilon\equiv P_H/\widehat P-1$ 
  of the model power spectrum $P_H$ around the average recovered best-fit value $\widehat P$. As the third order approximation is not normalisable, 
  the normalisation has been chosen such that it agrees with the 4$^{\text{th}}$ order at the maximum. The true posterior shape agrees very well with the inverse cubic normal
  posterior shape.}
 \label{fig:PRexpansion}
\end{figure}

We can see from Fig.~\ref{fig:PRexpansion} that the 3rd order diverges for large values of the model power spectrum $P_H$. Hence the optimal free parameter $a$ might differ from $a=-1/4$. Therefore, we use the KL divergence to optimise $a$ in the offset log-normal 
distribution $\posterior_{\rmn{OLN}}$. It can be found to be $a=-0.201$ at $k=0.004\Mpch$, $a=-0.240$ at $k=0.012\Mpch$ and $a=-0.242$ at higher values of $k$.
$\posterior_{\rmn{OLN}}$ peaks at the maximum of the true distribution $\posterior_R$ and it approximates the tails of the true distribution a bit better than the Gaussian approximations, 
but as Fig.~\ref{fig:OLN} shows, it is 
still obviously different from $\posterior_R$.

The ICN distribution \citep{Smith:2005ue} fits the true distribution better. Fig.~\ref{fig:PRexpansion} shows a remarkable agreement 
between $\posterior_R$ and $\posterior_{\rmn{ICN}}$. Writing both $-2\ln(\posterior_R)$ and $-2\ln(\posterior_{\rmn{ICN}})$ as Taylor series, we see that their Taylor coefficients are
equal for $k\leq 3$ and approximately equal for much higher orders (cf. Appendix~\ref{sec:Taylor}). 

\section{The Effect on $\lowercase{f}_\mathrm{NL}$ Measurements}
\label{sec:fNL}

\subsection{Physical Model}
\label{sec:fNLmodel}
In this section, we test the effect of using different posterior distribution shapes on the inference of a real observable. The largest deviations between the 
posteriors are at small $k$ and we would therefore expect the largest effects for parameters dependent on these modes. 
At these scales, (local) primordial non-Gaussianity alters the biasing law between dark-matter halos and the underlying mass-density field 
\citep{Dalal:2007cu, Matarrese:2008nc, Slosar:2008hx, Afshordi:2008ru, Valageas:2009vn, Giannantonio:2009ak, Schmidt:2010gw, Desjacques:2011jb}, 
making $f_\mathrm{NL}$
a perfect test parameter of our analysis. The parameter arises in models where the potential has a local quadratic term
\begin{equation}
 \Phi=\phi+f_\mathrm{NL}\left(\phi^2-\left\langle\phi^2\right\rangle\right).
\end{equation}
The resulting alteration of the bias can be written as
\begin{equation}
 b(k,f_\mathrm{NL})=b_0+\delta b(f_\mathrm{NL})+\Delta b(k,f_\mathrm{NL}),
 \label{eq:totalb}
\end{equation}
where $b_0$ is the bias in a Universe without primordial non-Gaussianity, $\delta b(f_\mathrm{NL})$ is the scale-independent modification to the bias from the non-Gaussian form
of the mass functions and \citep{Schmidt:2010gw, Desjacques:2011jb}
\begin{equation}
 \Delta b(k,f_\mathrm{NL})\approx(b_0-1)f_\mathrm{NL}A(k)
\end{equation}
is the local scale-dependent correction due to the easier halo formation with additional long-wavelength fluctuations, which depends on the critical density $\delta_c(z)$ in the peak-
background split model, as well as the matter transfer function $T(k)$, the matter density $\Omega_m$, the present-time Hubble parameter $H_0$ and the linear growth function $D(z)$
through the parameter
\begin{equation}
 A(k,z)=\frac{3\Omega_m\delta_c(z)}{k^2T(k)}\left(\frac{H_0}{c}\right)^2.
\end{equation}
As $\delta b(f_\mathrm{NL})\ll \Delta b(k,f_\mathrm{NL})$
at our scales of interest \citep{Slosar:2008hx, Giannantonio:2009ak}, we neglect $\delta b(f_\mathrm{NL})$. 

\begin{figure}
 \centering
 \includegraphics[width=\linewidth]{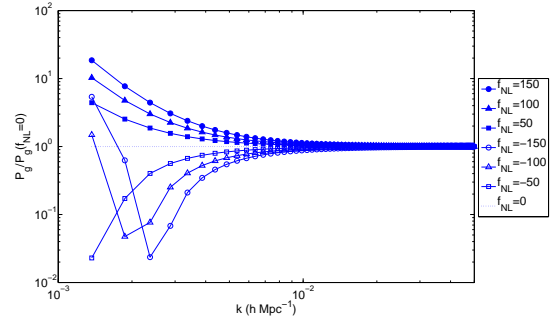}
 \caption{Galaxy power spectra $P_g$ calculated for different values of $f_\mathrm{NL}$ divided by the galaxy power spectrum $P_g(f_{NL}=0)$ of a universe with a Gaussian primordial density field.}
 \label{fig:PoverPfNL0}
\end{figure}
Fig.~\ref{fig:PoverPfNL0} shows the effect of $f_\mathrm{NL}$ on the galaxy power spectrum at large scales. We plot the galaxy power spectrum $P_g$ divided by the galaxy power spectrum at $f_\mathrm{NL}=0$, hence what we plot is proportional to the square of Eq.~\eqref{eq:totalb}. At lowest $\k$, negative $f_\mathrm{NL}$ enhances the power spectrum due to the fact that the term proportional to $f_\mathrm{NL}^2$ dominates the total bias. At slightly higher $\k$, but still at large scales, the term linear in $f_\mathrm{NL}$ dominates, and the power is enhanced or decreased depending on the sign of $f_\mathrm{NL}$. At small, yet still linear, scales, $A(k,z)$ becomes small, thus initial local non-Gaussianities do not have an effect on the galaxy power spectrum at these scales.

Here we work to first order in $\delta$, so that we can continue to assume that $\delta_\k$ is drawn from a Gaussian distribution, 
with an altered $P(\k)$, i.e. the first order effect of non-Gaussianity is to $P_H(\k)$, keeping the distribution the same. Furthermore, we do not alter $V_n$
(Eq.~\eqref{eq:Vn}) to include any coupling between modes from the non-Gaussian signal. Where $k$ is very small, higher order corrections to $\delta$ will become 
increasingly important 
\citep[e.g.][]{Tellarini:2015faa}, suggesting that the Gaussian limit for $\delta$ will break down here.

\subsection{BOSS Results}
\label{sec:fNLBOSS}
\begin{figure}
 \centering
 \includegraphics[width=\linewidth]{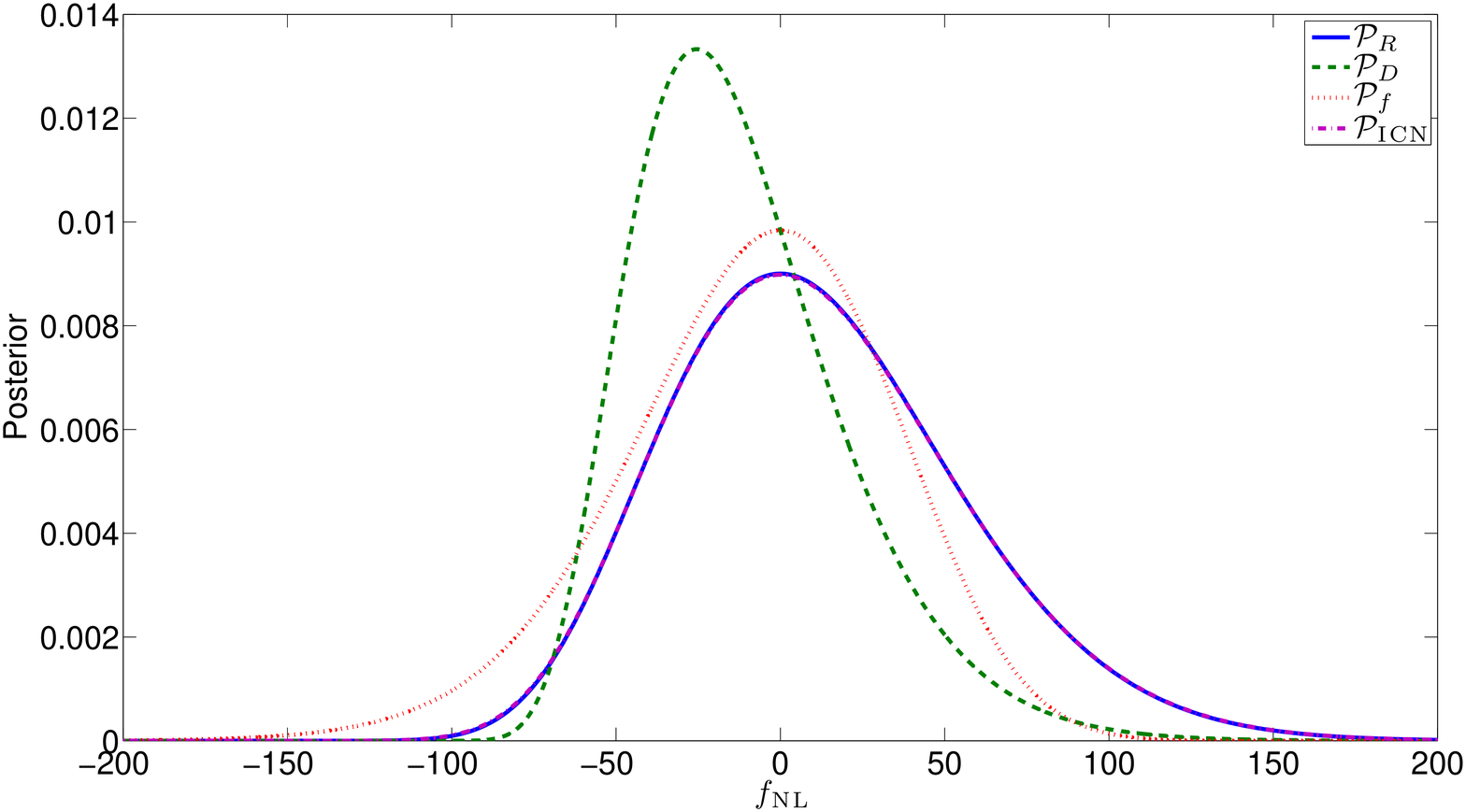}
 \caption{Analytic $f_\mathrm{NL}$-posterior functions for a BOSS like survey combining all $k$-bins.}
 \label{fig:BOSSfNLpost}
\end{figure}

We use BOSS DR9 parameters and the same CAMB linear matter power spectrum as \citet{Ross:2012sx}. We also assume $\delta_c=\frac{1.686}{D(z)}$ 
as expected from the spherical collapse model in an Einstein-de Sitter
universe and a flat prior for $f_\mathrm{NL}$. We plot $f_\mathrm{NL}$ posterior functions in Fig.~\ref{fig:BOSSfNLpost}, 
assuming a measurement of a power spectrum with underlying $f_\mathrm{NL}=0$. $\posterior_f$ is not symmetric, as both a linear 
and a quadratic term of $f_\mathrm{NL}$ enter the power spectrum. The inverse cubic normal distribution agrees again
very well with $\posterior_R$. 
$\posterior_R$, $\posterior_f$ and $\posterior_\mathrm{ICN}$ reproduce the true value as their best fit estimate. Using $\posterior_D$, the most likely value of 
$f_\mathrm{NL}$ is $f_\mathrm{NL}=-25.5$
considering the same $k$-bins as \citet{Ross:2012sx} in their analysis of DR9 BOSS data, i.e. 
$0.004\frac{h}{\mathrm{Mpc}}\leq k\leq 0.05\frac{h}{\mathrm{Mpc}}$. 

One has to keep in mind that there are different definitions of the measured value. The commonly published value is the posterior mean $\langle f_\mathrm{NL}\rangle$, due to
the fact that
if $f_\mathrm{NL}$ is fitted as part of a longer list of cosmological parameters, one has to rely on Markov chain Monte Carlo techniques \citep[e.g.][]{Lewis:2002ah}. 
In general, such techniques cannot provide accurate estimates of the best-fit value. Hence, data analysis papers more often present $\langle f_\mathrm{NL}\rangle$ 
as their results. If the posterior is asymmetric, the best fit and posterior mean do not agree.
Given a flat $f_\mathrm{NL}$-prior, we expect $f_\mathrm{NL}=11.4$ using $\posterior_R$. Based on our arguments in Sec.~\ref{sec:deltakL} and \ref{sec:fNLmodel}, we think of the mean
of $\posterior_R$ as the correct estimate of $f_\mathrm{NL}$. This seems counter-intuitive because our input was that we measure a power spectrum which corresponds to 
$f_\mathrm{NL}=0$, but we have to consider that $\Phat$ is a finite empirical realisation in our part of the universe corresponding to the value of $f_\mathrm{NL}=0$ we have assumed we would measure locally, but due to the non-Gaussian shape of the
posterior distribution, the ensemble average of $f_\mathrm{NL}$ measured in other parts of the universe is higher than the value we set as an input for our local environment.

Our results are summarised in Tab.~\ref{tab:DR9fNLpostdict}. $\posterior_\mathrm{ICN}$ reproduces the correct estimate of $f_\mathrm{NL}$, whereas $\posterior_D$ and $\posterior_f$ 
estimate $f_\mathrm{NL}=-11.9$ and $f_\mathrm{NL}=-7.7$ respectively. The choice of the posterior distribution also affects the error estimation. If we use $\posterior_R$ or 
$\posterior_\mathrm{ICN}$, 
the length of our postdicted 95\% $f_\mathrm{NL}$-confidence interval (C.I., cf. Tab.~\ref{tab:DR9fNLpostdict})
is similar to the length of \citet{Ross:2012sx}'s
most na\"ive case ii 95\% C.I., i.e. $32<f_\mathrm{NL}<198$.

\begin{table}
 \centering
 \begin{minipage}{\linewidth}
  \caption{$f_\mathrm{NL}$-postdictions of the best fit $f_\mathrm{NL}^\mathrm{(BF)}$ and marginalised best fits $\left\langle f_\mathrm{NL}\right\rangle$, 
  as well as its 95\% confidence interval, for BOSS DR9 using different 
  shapes of the posterior distribution.}
  \label{tab:DR9fNLpostdict}
  \begin{tabular}{@{}lrrl@{}}
  \hline
   posterior & $f_\mathrm{NL}^\mathrm{(BF)}$ & $\langle f_\mathrm{NL}\rangle$ &
   95\% confidence interval \\
 \hline
  $\posterior_R$ & 0 & 11.4 & -71.5$<f_\mathrm{NL}<$100.7 \\
  $\posterior_D$ & -25.5 & -11.9 & -68.2$<f_\mathrm{NL}<$53.4 \\
  $\posterior_f$ & 0 & -7.7 & -90.9$<f_\mathrm{NL}<$71.0 \\
  $\posterior_\mathrm{ICN}$ & 0 & 11.4 & -71.9$<f_\mathrm{NL}<$101.2 \\
\hline
\end{tabular}
\end{minipage}
\end{table}

\subsection{Euclid Results}
\label{sec:fNLEuclid}
We make similar predictions for the Euclid survey \citep{Laureijs:2011gra}. We assume bias values $b(z)=\sqrt{1+z}$, matched to simulations of \citet{Orsi:2009mj} and also
assumed in \citet{Amendola:2012ys}, and number densities $\nbar(z)$ 
predicted for Euclid by \citet*{Pozzetti}, and a survey covering 15000 square degrees. We generate CAMB matter power spectra $P(\k, z)$ for the redshift range $0.9<z<1.74$. 
Note that the aim of this article is to test how the use of different 
posterior shapes influences cosmological measurements, but not primarily to make $f_\mathrm{NL}$-predictions. We refer to more 
rigorous predictions which can be found e.g. in \citet*{Fedeli:2010ud, Laureijs:2011gra, Giannantonio:2011ya, Yamauchi:2014ioa}.
These studies also include 3-point statistics, 
weak lensing tomography, measurements of the integrated 
Sachs-Wolfe effect and/or the use of the multitracer technique. Their constraints are therefore tighter than ours.

\begin{figure}
 \centering
 \includegraphics[width=\linewidth]{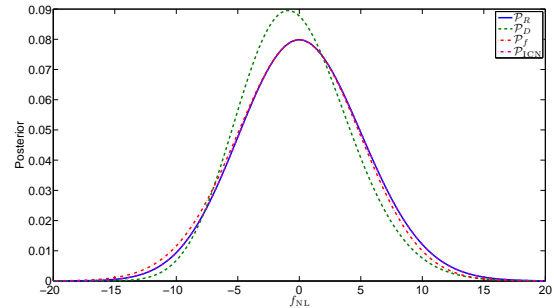}
 \caption{Analytic $f_\mathrm{NL}$-posterior functions for a Euclid like survey combining all $k$-bins.}
 \label{fig:EuclidfNLpost}
\end{figure}

As Euclid will probe a much larger volume, it will accommodate many more $k$-modes and hence we see good agreement of $\posterior_f$ with $\posterior_R$ in 
Fig.~\ref{fig:EuclidfNLpost}. As against our results in Sec.~\ref{sec:BOSS}, 
fixing the covariance provides better $f_\mathrm{NL}$ results than the inferences from a posterior with varying covariance. 
However, $\posterior_{\mathrm{ICN}}$ is still the best approximation and 
accurately reproduces the marginalised $f_\mathrm{NL}$-value of $\posterior_R$ and its 95\% C.I., whereas using $\posterior_f$ yields the correct width of the 95\% C.I., but its 
position and the marginalised value have an offset of 0.38 (cf. Tab.~\ref{tab:EuclidfNLpostdict}). We therefore still recommend either using $\posterior_\mathrm{ICN}$ or $\posterior_R$ when cosmological
models are fitted to power spectra from galaxy surveys even as large as Euclid.

\begin{table}
 \centering
 \begin{minipage}{\linewidth}
  \caption{$f_\mathrm{NL}$-predictions similar to Tab.~\ref{tab:DR9fNLpostdict}, but for Euclid.}
  \label{tab:EuclidfNLpostdict}
  \begin{tabular}{@{}lrrl@{}}
  \hline
   posterior & $f_\mathrm{NL}^\mathrm{(BF)}$ & $\left\langle f_\mathrm{NL}\right\rangle$ &
   95\% confidence interval \\
 \hline
  $\posterior_R$ & 0 & 0.24 & -9.0$<f_\mathrm{NL}<$9.4 \\
  $\posterior_D$ & -1.0 & -0.30 & -8.4$<f_\mathrm{NL}<$7.8 \\
  $\posterior_f$ & 0 & -0.14 & -9.4$<f_\mathrm{NL}<$9.0 \\
  $\posterior_\mathrm{ICN}$ & 0 & 0.24 & -9.0$<f_\mathrm{NL}<$9.4 \\
\hline
\end{tabular}
\end{minipage}
\end{table}

\section{Conclusions}
\label{sec:conclusion}

We have studied different posterior shapes that can be used in the fitting process of cosmological models to power spectra from galaxy surveys. As the underlying matter density field
is at least approximately Gaussian, we assume that the true posterior distribution $\posterior_R$ 
is based on a Rayleigh likelihood distribution in $\delta$. Assuming Gaussian posteriors in $P(\k)$, be it with a fixed
or a varying covariance matrix, does not approximate $\posterior_R$ well and yields biased best-fit values and wrong error estimates especially on large scales where statistics are
not good enough to make use of the central limit theorem. 

If one confines oneself to use Gaussian posterior shapes, it 
depends on the parameter one wants to constrain whether a fixed or varying covariance matrix provides more accurate results. We found that the posterior shape $\posterior_D$ with
varying covariance follows $\posterior_R$ closer than $\posterior_f$ with a fixed covariance when the power spectrum $P_H$ (or any parameter linear in the power
spectrum) is fitted to the power spectrum $\widehat P$, but when $f_\mathrm{NL}$ is fitted to $\widehat P$ it is the other way round. 

Due to these reasons, we advise against using Gaussian posterior distributions. Instead, we have found that posterior distributions, such as the inverse cubic normal distribution $\posterior_\mathrm{ICN}$ (cf. Eq.~\eqref{eq:PICN}) or applying \citet{Hamimeche:2008ai}'s method to $\posterior_R$ (cf. Eq.~\eqref{eq:PHL}),
provide simple, more accurate alternatives. They confidently reproduce the correct width of the 95 \% confidence intervals in our simplified predictions of $f_\mathrm{NL}$-measurements. However, the final decision about which posterior is the best to use should be done after testing these 
methods against simulations which account for the non-linear effects that we have ignored for simplicity in our analytic calculations. We leave this for future work.

A major advantage of the non-Gaussian posteriors presented in this paper, is the fact that 
their covariance matrices do not depend on the power spectrum of the model to be tested. The estimation of covariance matrices is a critical and 
computationally expensive step in the data analysis. Extensions to configuration-space analyses based on the correlation function $\xi(\mathbf{r})$ are left for future work.

\section*{Acknowledgments}

The authors would like to thank Matteo Tellarini, Ashley Ross and David Wands for valuable discussions about primordial non-Gaussianity. We thank the referee Andrew Jaffe for
his helpful comments.

Some of the results in this paper have been generated using the CAMB package \citep{Lewis:1999bs}. For some of the other results, we made use of the facilities and staff
of the UK Sciama High Performance Computing cluster supported
by the ICG, SEPNet and the University of Portsmouth.

WJP acknowledges support from UK STFC through the consolidated grant ST/K0090X/1, and from the European Research
Council through the Darksurvey grant. LS is grateful for support from SNSF grant SCOPES IZ73Z0-152581, GNSF grant FR/339/6-350/14,
and DOE grant DEFG 03-99EP41093.

\appendix

\section{Comparison of $\posterior_R$ and $\posterior_{\rmn{ICN}}$ Taylor Series}
\label{sec:Taylor}

In this appendix, we compare the Taylor Series of $\posterior_R$ and $\posterior_{\rmn{ICN}}$ to explain why they are so similar.
We write the hypothetical power spectrum $P_H\equiv (1+\epsilon)\widehat{\vert\delta_\k\vert}^2$ as a perturbation around the measured power. The Rayleigh posterior hence becomes 
\begin{equation}
 -2\ln(\posterior_R)=2\ln(1+\epsilon)+\frac{2}{1+\epsilon}.
\end{equation}
Ignoring the irrelevant zero order contribution, the Taylor series reads
\begin{equation}
 -2\ln(\posterior_R)=2\sum_{\kappa=1}^\infty (-1)^\kappa\epsilon^\kappa\frac{\kappa-1}{\kappa}.
\end{equation}
The ICN distribution in terms of $\epsilon$ is given by
\begin{align}
 -2\ln(\posterior_{\rmn{ICN}})&=9\left[1-(1+\epsilon)^{-1/3}\right]^2\nonumber\\
  &=9\left[1-2(1+\epsilon)^{-1/3}+(1+\epsilon)^{-2/3}\right].
\end{align}
We make use of the generalised binomial series $(1+\epsilon)^\alpha=\sum_{\kappa=0}^\infty\binom{\alpha}{\kappa}\epsilon^\kappa$, where 
$\binom{\alpha}{\kappa}\equiv\frac{\Gamma(\alpha+1)}{\Gamma(\kappa+1)\Gamma(\alpha-\kappa+1}$ is the generalised binomial coefficient, and obtain the series
\begin{equation}
 -2\ln(\posterior_{\rmn{ICN}})=9\sum_{\kappa=1}^\infty \epsilon^\kappa\left[\binom{-\frac{2}{3}}{\kappa}-2\binom{-\frac{1}{3}}{\kappa}\right].
\end{equation}
Again, we have ignored irrelevant constant terms. The negative entries in the binomial coefficients can be removed using 
$\binom{\alpha}{\kappa}=(-1)^\kappa\binom{\kappa-\alpha-1}{\kappa}$:
\begin{equation}
 -2\ln(\posterior_{\rmn{ICN}})=9\sum_{\kappa=1}^\infty (-1)^\kappa\epsilon^\kappa\left[\binom{\kappa-\frac{1}{3}}{\kappa}-2\binom{\kappa-\frac{2}{3}}{\kappa}\right].
\end{equation}
If we insert values for $\kappa\leq 3$, we find the equality 
\begin{equation}
 2\frac{\kappa-1}{\kappa}=9\left[\binom{\kappa-\frac{1}{3}}{\kappa}-2\binom{\kappa-\frac{2}{3}}{\kappa}\right].
\end{equation}
Thus, $\posterior_R$ and $\posterior_{\rmn{ICN}}$ are the same to third order. What is even more striking is that for larger $\kappa$,the approximation
\begin{equation}
 2\frac{\kappa-1}{\kappa}\approx 9\left[\binom{\kappa-\frac{1}{3}}{\kappa}-2\binom{\kappa-\frac{2}{3}}{\kappa}\right].
\end{equation}
still holds. For $\kappa<17$, the two sides differ by less than 20\%.
Therefore, the agreement between $\posterior_R$ and $\posterior_{\rmn{ICN}}$ is high. 

\bsp

\label{lastpage}

\end{document}